# Disordered one-dimensional photonic structures composed by more than two materials with the same optical thickness

Ilka Kriegel[1], Francesco Scotognella[1,2]*

[1] Dipartimento di Fisica, Politecnico di Milano, Piazza Leonardo da Vinci 32, 20133 Milano
[2] Istituto di Fotonica e Nanotecnologie CNR, Piazza Leonardo da Vinci 32, 20133 Milano
* francesco.scotognella@polimi.it

**Abstract**
The optical properties of photonic structures made with more than two materials are very interesting for optical filtering and lighting applications. Herein, we compared the transmission properties of one-dimensional photonic crystals made with three, four and five materials, showing that, with a photonic crystal made of $t$ different materials, the band gap splits in $t$-1 bands. The same optical thickness for the different materials layers results in split photonic band gaps with the same intensity. Photonic crystals with more than two materials result in a simple structure that could be used for multi-feature optical filters, or that could provide feedback for multi-emission distributed feedback lasers. Furthermore, we analyzed the transmission properties of ternary and quaternary random photonic structures. These materials could very interesting for light trapping in photovoltaic devices.

**Introduction**
Photonic crystals provide the possibility to manipulate light: a periodic modulation of the refractive index results in a forbidden energy gap that excludes the existence of optical modes within a certain range of wavelengths [1-6]. In the last twenty years, several methods have been established to fabricate photonic crystals at diverse length scales. as layer-by-layer stacking techniques using microfabrication tools [7–9], electrochemical etching [10,11], laser-beam-scanning chemical vapour deposition [12], and holographic lithography [13,14].
The periodicity of dielectric constant can be in one, two and three dimensions [15-18]. In the case of one-dimensional structures, well established and low-cost fabrication technologies, as spin coating or co-extrusion, can be employed [15,19]. Recently, these materials are extensively studied since they find application in several fields, including photonics for low threshold laser action, high bending angle waveguide, super-prism effect, sensors and optical switches [20-25]. The calculation of the optical properties can be analytically performed with several theoretical methods, as a fully-vectorial algorithm to compute the definite-frequency eigenstates of Maxwell's equations, developed by Johnson and Joannopoulos [26], or the transfer matrix method [27]. In particular, the transfer matrix approach also allows to study the properties of photonic quasicrystals [28] and disordered photonic structures [29]. In the case on one-dimensional disordered, it is possible to employ also a finite element method [30]. Owing to their interesting features, the study of one-dimensional photonic crystals (1DPCs) made with more than two materials is recently increased. For example, ternary photonic crystals have been theoretically investigated [31,32]. With these crystals it is usually shown omnidirectional reflectance; an interesting case is a ternary Fibonacci sequence in which one of the three materials is a superconductor [33]. In 2012 quaternary photonic crystals have been proposed [34].
In this work, we compared the transmission properties of one-dimensional photonic crystals made with three, four and five materials, showing that, with a photonic crystal made of $t$ different materials, the band gap splits in $t$-1 bands. Furthermore, we analyzed the total light transmission of ternary and quaternary random photonic structures as a function of the refractive index contrast. We observed that the total transmission of the periodic photonic crystals and the one of the disordered structures show a different behaviour.

**Methods**
The photonic structures here realized are multilayers, in which the layers have refractive indexes $n_i$ and thicknesses $(300\ nm)/n_i$. In this way, all the layers have the same optical thickness. The periodic photonic crystals made with two, three, four and five different materials have been engineered with a unit cell characterized by a monotonic increase of the refractive index. For example, in the quaternary photonic crystal the unit cell is made with $n_1$, $n_2$, $n_3$ and $n_4$, where $n_1 < n_2 < n_3 < n_4$.

The random ternary photonic structure have been realized by giving to each layer the 33.3% to have refractive index $n_1$, $n_2$ or $n_3$ (as flipping a hypothetical coin with three faces). The random quaternary photonic structure have been realized by giving to each layer the 25% to have refractive index $n_1$, $n_2$, $n_3$ or $n_4$ (as flipping a dice with four faces).

Calculations of the light transmission through the media (in the spectral region corresponding to the photonic band gap of the periodic photonic crystal) are performed with the transfer matrix method [27]. We have considered isotropic, non-magnetic materials shaping the system glass/multilayer/air (in which glass is the sample substrate) and an incidence of the light normal to the stacked layer surface. $n_0$ and $n_S$ are the refractive indexes of air and glass, respectively, while $E_m$ and $H_m$ are the electric and magnetic fields in the glass substrate. To determine the electric and magnetic fields in air, $E_0$ and $H_0$, we have solved the following system:

$$\begin{bmatrix} E_0 \\ H_0 \end{bmatrix} = M_1 \cdot M_2 \cdot \ldots \cdot M_m \begin{bmatrix} E_m \\ H_m \end{bmatrix} = \begin{bmatrix} m_{11} & m_{12} \\ m_{21} & m_{22} \end{bmatrix} \begin{bmatrix} E_m \\ H_m \end{bmatrix} \quad (1)$$

where

$$M_j = \begin{bmatrix} A_j & B_j \\ C_j & D_j \end{bmatrix},$$

with $j=(1,2,\ldots,m)$, is the characteristic matrix of each layer. The elements of the transmission matrix $ABCD$ are

$$A_j = D_j = \cos(\phi_j),\ B_j = -\frac{i}{p_j}\sin(\phi_j),\ C_j = -i\,p_j \sin(\phi_j), \quad (2)$$

where $n_j$ and $d_j$, contained in the angle $\phi_j$, are respectively the effective refractive index and the thickness of the layer $j$. In the case of normal incidence of the light beam, the phase variation of the wave passing the $j$-fold layer is $\phi_j = (2\pi/\lambda)n_j d_j$, while the coefficient $p_j = \sqrt{\varepsilon_j/\mu_j}$ in transverse electric (TE) wave and $q_j=1/p_j$ replace $p_j$ in transverse magnetic (TM) wave. Inserting Equation (2) into Equation (1) and using the definition of transmission coefficient,

$$t = \frac{2 p_s}{(m_{11} + m_{12} p_0) p_s + (m_{21} + m_{22} p_0)} \quad (3)$$

it is possible to write the light transmission as

$$T = \frac{p_0}{p_s}|t|^2 \quad (4)$$

The transmission spectra are calculated as a function of the photon energy, with steps of 2 meV.

**Results and Discussion**

We have calculated the transmission spectra of photonic crystals with unit cells composed by two, three, four and five materials. The spectra are shown in Figure 1. The common two-material photonic crystal show the fundamental band gap at about 1 eV and the second order photonic band gap at 3 eV circa. The photonic crystals with more than two materials show a splitting of the photonic band gaps. For the three-material one, each gap is splitting in two gaps. The same is occurring for the other photonic crystals, resulting in this observations: the photonic crystals made with $t$ different materials show a photonic band gaps split in $t$-1 bands.

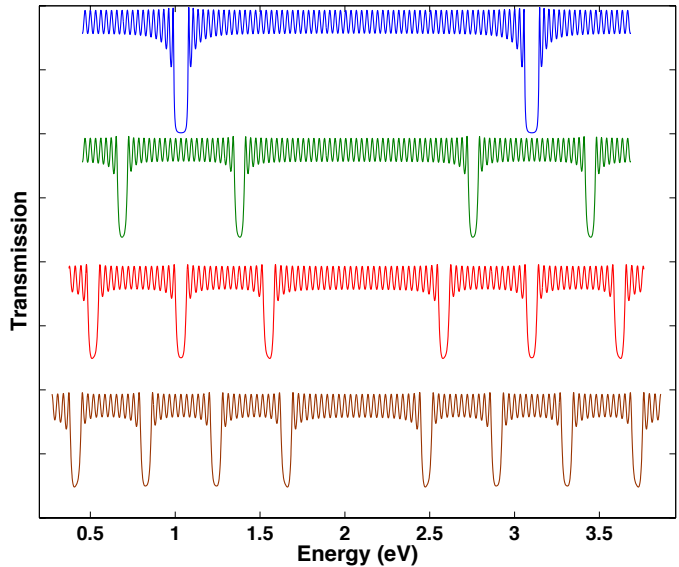

**Figure 1.** Transmission spectra of a photonic crystal (blue curve), a ternary photonic crystal (green curve), a quaternary photonic crystal (red curve) and a five material photonic crystal (brown curve).

These types of photonic crystals could be very interesting for the fabrication of optical filters, with many rejection bands. It is worth noting that the split gaps have the same intensity. We want to stress that this happens when all the layers have the same optical thickness. For example, the two-material photonic crystals have layers with refractive indexes $n_1$ and $n_2$. According to the Bragg-Snell law, the wavelength corresponding to the band gap minimum is $\lambda_B$ = 2($n_1 d_1 + n_2 d_2$) = 2($n_1(300/n_1) + n_2(300/n_2)$) = 1200 nm (1.03 eV), for any choice of $n_1$ and $n_2$. The aforementioned scenario is occurring with a periodic sequence of unit cells. It is interesting to compare the periodic crystals with disordered structures. In Figure 2 we compare the transmission spectra of three-material periodic photonic crystals with the one of three-material random photonic structures. In the spectral region of the split photonic band gaps (for the periodic crystal) the disordered structure shows many transmission depths.

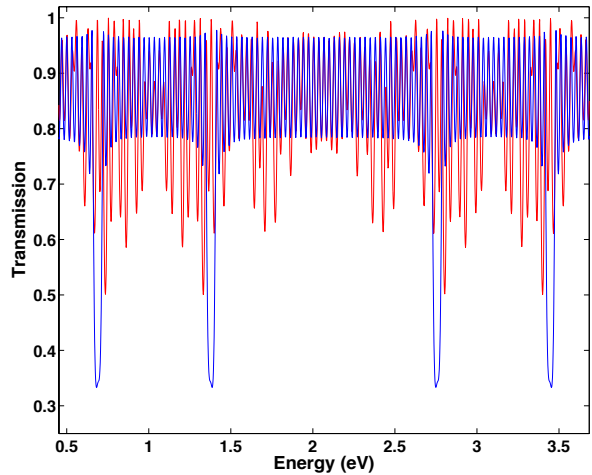

a)

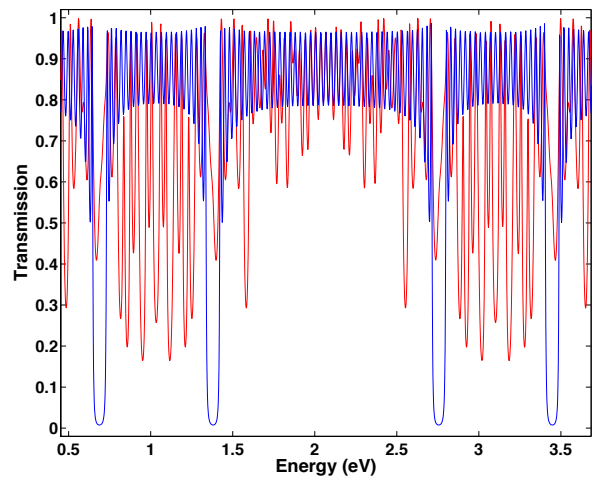

b)

**Figure 2.** a) transmission spectrum of a ternary photonic crystal (blue curve) and random ternary photonic structure (red curve) with refractive indexes $n_1$=1.92, $n_2$=2, $n_3$=2.08; b) transmission spectrum of a ternary photonic crystal (blue curve) and random ternary photonic structure (red curve) with refractive indexes $n_1$=1.8, $n_2$=2, $n_3$=2.2.

The intensity of the transmission depths is more intense in the case of higher refractive index contrast between the layers (Figure 2b). This phenomenon is even more clear for four-material structures (Figure 3a and Figure 3b). It is worth noting that, for both the ternary and quaternary random structures, and for all the refractive index contrast, the more intense transmission depths are concentrated in the spectral region where the periodic crystals show the photonic band gaps.

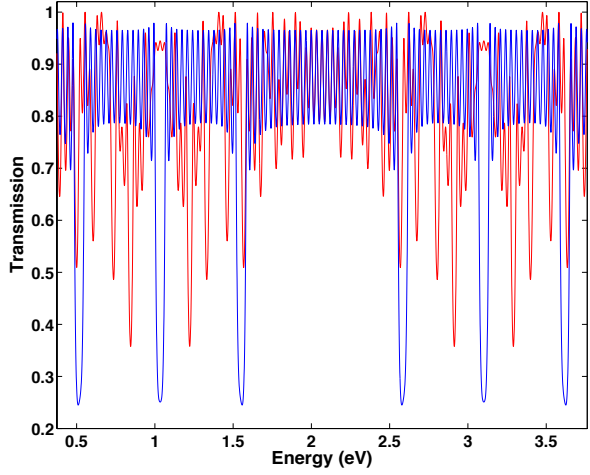

a)

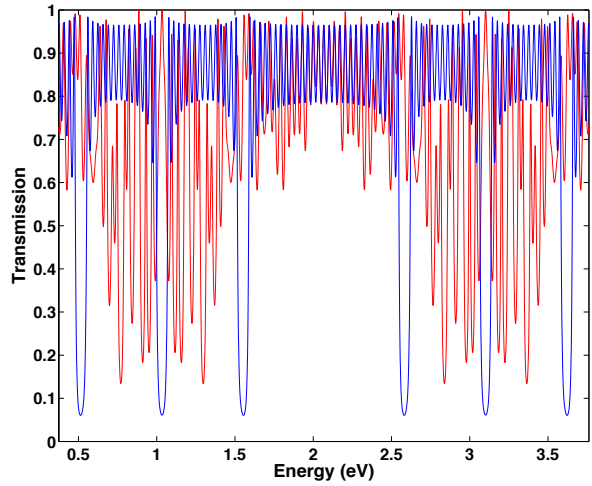

b)

**Figure 3.** a) transmission spectrum of a quaternary photonic crystal (blue curve) and random quaternary photonic structure (red curve) with refractive indexes $n_1$=1.85, $n_2$=1.95, $n_3$=2.05, $n_4$=2.15; b) transmission spectrum of a quaternary photonic crystal (blue curve) and random quaternary photonic structure (red curve) with refractive indexes $n_1$=1.775, $n_2$=1.925, $n_3$=2.075, $n_4$=2.225.

We then consider the total transmission of the structures in the region of the fundamental and second order gaps (with a step of 2 meV). This zone includes two photonic band gaps regions and a transparent region. In Figure 3 we plot the overall transmission as a function of $n_{max}$-$n_{mean}$, where $n_{max}$ is the highest refractive index in the structure, while $n_{mean}$ is its mean refractive index. For the disordered structures, each overall transmission (for a certain $n_{max}$-$n_{mean}$) is a average value for 1000 permutations of layer sequences. In the case of three-material structures (Figure 4a) we observed that, for a lower refractive index contrast, the overall transmission is higher for random structures, while for a higher refractive contrast the overall transmission is higher for periodic crystals. Instead, in the case of four-material structures, the overall transmission is always higher for periodic crystals. In Table 1 and Table 2 we report the values in Figure 4a and 4b, respectively.

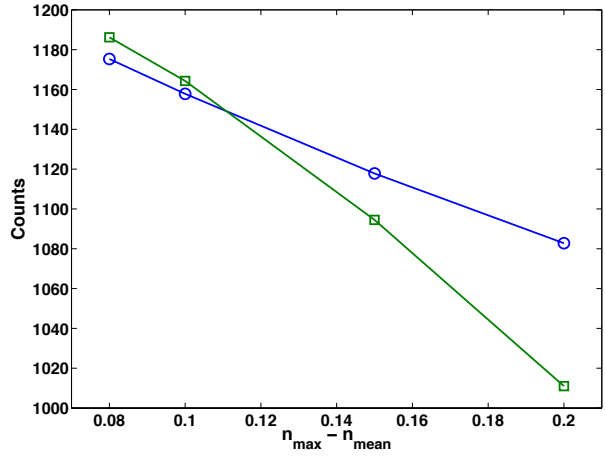

a)

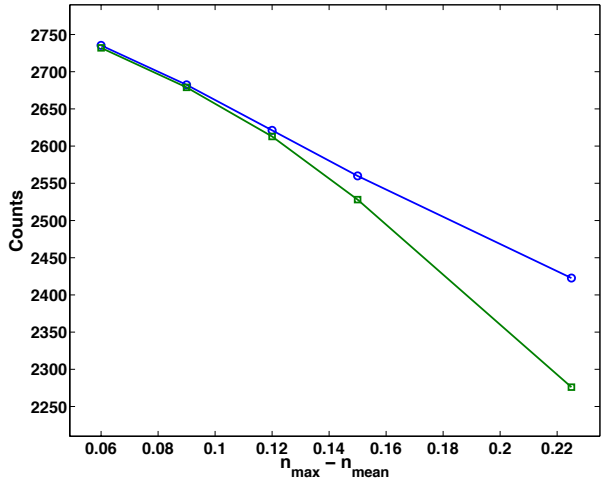

b)

**Figure 4.** a) Average total transmission for the ternary photonic crystal (blue curve) and the random ternary photonic structure as a function of $n_{max}$-$n_{mean}$; b) average total transmission for the quaternary photonic crystal (blue curve) and the random quaternary photonic structure as a function of $n_{max}$-$n_{mean}$.

**Table 1.** Total transmission for the ternary photonic structures, as reported in Figure 4a. The values for the disordered structures are averaged over 1000 permutations of layer sequences.

| $n_{max}$-$n_{mean}$ | Total T ordered | Total T disordered |
|---|---|---|
| 0.08 | 1175.3 | 1186.2 |
| 0.1 | 1157.8 | 1164.3 |
| 0.15 | 1117.8 | 1094.5 |
| 0.2 | 1082.8 | 1011.0 |

**Table 2.** Total transmission for the quaternary photonic structures, as reported in Figure 4a. The values for the disordered structures are averaged over 1000 permutations of layer sequences.

| $n_{max}$-$n_{mean}$ | Total T ordered | Total T disordered |
|---|---|---|
| 0.06 | 2735.4 | 2732.1 |
| 0.09 | 2682.4 | 2679.0 |
| 0.12 | 2621.2 | 2612.9 |
| 0.15 | 2559.9 | 2528.1 |
| 0.225 | 2422.6 | 2276.0 |

Finally, we have analysed one of the two split photonic band gaps of the three material photonic crystal, for all the possible permutations of the three different materials in the unit cell. The photonic band gap is only slightly different, but and effect is stronger in the side bands.

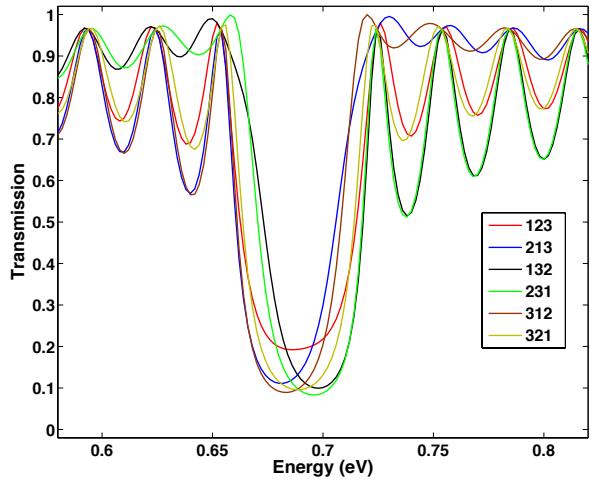

**Figure 5.** Transmission spectra of three-material photonic crystals with the six different permutations in the unit cell of three materials ($n_1$=1.9, $n_2$=2, $n_3$=2.1).

## Conclusions

In this work, we compared the transmission properties of one-dimensional photonic crystals made with three, four and five materials, showing that, with a photonic crystal made of $t$ different materials, the band gap splits in $t$-1 bands. Each layer employed to make the photonic structures has the same optical thickness, and this results in split photonic band gaps with the same intensity. Photonic crystals with more than two materials result in a simple structure that could be used for multi-feature optical filters, or that could provide feedback for multi-emission distributed feedback lasers. Furthermore, we analyzed the total transmission of ternary and quaternary random photonic structures, and we observed that such total transmission, in the studied spectral region, is lower for disordered structures. This happens for a refractive index contrast ≥0.15 in the case of three-material structures, and for any refractive index contrast for four-material structures. These materials could very interesting for light trapping in photovoltaic devices.